\documentclass[%
 reprint,
 amsmath,amssymb,
 aps,
 prl,
 longbibliography,
 lengthcheck,%
]{revtex4-1}

\usepackage{graphicx}
\usepackage{dcolumn}
\usepackage{bm}
\usepackage{hyperref}

\begin{document}

\preprint{APS/123-QED}

\title{Quantum kinetics of spinning neutral particles: General theory and Spin wave dispersion}

\author{P. A. Andreev}%
\email{andreevpa@physics.msu.ru}
\affiliation{
 Department of General Physics,
 Faculty of Physics, Moscow State
University, Moscow, Russian Federation.}

\date{\today}

\begin{abstract}
Plasma physics give an example of physical system of particles with the long range interaction. At small velocity of particles we can consider the plasma approximately as a system of particles with the Coulomb interaction. The Coulomb interaction is isotropic. Systems of spinning neutral particles have long-range anisotropic interparticle interaction. So, they can reveal more reach properties than plasma. Furthermore for studying of systems of spinning particles we can develop kinetic and hydrodynamic methods analogous to used for the plasma.
We derive kinetic equations by a new method, which is the generalization of the many-particle quantum hydrodynamics. Obtained set of kinetic equations is truncated, so we have closed set of two equations. One of them is the kinetic equation for quantum distribution function. The second equation is the equation for the spin-distribution. Which describes the spin kinetic evolution and gives contribution in time evolution of the distribution function. Our method allows to obtain equations as for three dimensional system of particles and for low dimensional systems. So, we consider spin waves in three- and two dimensional systems of neutral spinning particles.
 \end{abstract}


\maketitle


\section{I. Introduction}

Studying of the quantum plasmas involves consideration of the spin evolution and its influence on the degenerate plasma properties. Corresponding hydrodynamic and kinetic equations are required for the quantum plasma studying. Method of direct derivation of the quantum hydrodynamic equations from the many-particle Schrodinger equation was suggested in 1999-2001 \cite{MaksimovTMP 1999}, \cite{MaksimovTMP 2001}, \cite{MaksimovTMP 2001 b}. Spin evolution and its contribution in the set of quantum hydrodynamic equations were considered there along with the exchange interactions. Features of exchange interaction contribution in the quantum hydrodynamic equations for bosons and fermions were demonstrated in these papers. These general properties were also used for the Bose-Einstein condensate and degenerate Fermi gas of neutral atoms \cite{Andreev PRA08}, where exchange interaction plays main role. The Wigner quantum kinetic was also used for quantum plasma description and getting of the corresponding quantum hydrodynamic equations \cite{Manfredi PRB 01}. A lot of papers have been published since when. Some results were discussed in review \cite{Shukla RMP 11}, \cite{P.K. Shukla UFN 10}. In this paper we suggest a new method of quantum kinetic equations derivation. We apply derived equations to study of waves in a system of neutral spinning particles, which gives an example of anisotropic long-range interaction. In this system the Coulomb interaction does not obscure the spin-spin interaction.

We present a method of quantum kinetic equation derivation.
This method is the direct generalization of basic ideas of the many-particle quantum hydrodynamics suggested
by L.S. Kuz'menkov and S.G. Maksimov \cite{MaksimovTMP 1999}, \cite{MaksimovTMP 2001}. We use an operator of quantum distribution function of $N$ particles, which corresponds to the classical one.
Our definition of the quantum kinetic function also corresponds to the explicit definition of particle concentration, which one of the main definition in the many-particle quantum hydrodynamics \cite{MaksimovTMP 1999}.
Considering spinning particles with the spin-spin interaction we come to a set of two kinetic equations, one for the distribution function, and another one for the spin distribution function. Both these functions have three arguments $f=f(\textbf{r},\textbf{p},t)$ and $\textbf{S}=\textbf{S}(\textbf{r},\textbf{p},t)$.
Presented here theory is suitable for different physical systems \cite{Andreev kinetics 12}. First of all it can be applied to the quantum plasma, where one studies evolution of charged spinning particles. However we focus our attention on neutral spinning particles to study the spin waves in magnetized dielectrics. A distribution function depending on four arguments has been considered in literature \cite{Brodin PRL 08 a}, \cite{Asenjo PL A 09}, \cite{Asenjo kinetic 12}. These arguments are time, space coordinate, momentum, and spin. In this case, a dependence of an equilibrium distribution function on spin has been considered (see Ref. \cite{Asenjo PL A 09} formula (2)). This description is based on the Wigner's theory \cite{Wigner PR 84}. There are other assumptions for quantum kinetics along with the Wigner's treatment \cite{Andreev kinetics 12}, \cite{Maksimov TMP 2002}, \cite{Maksimov P D 09}, \cite{Kuzelev PU 99}, \cite{Vagin 09}.

It is well-known that the Pauli equation for one charged spinning particle in an external electromagnetic field can be presented in a form of hydrodynamic equations \cite{Takabayasi PTP 55}. They are the continuity, the Euler (equation of the velocity field evolution), the Bloch (equation for spin density evolution) equations. The Euler equation for one particle contains density of force an external field acts on the particle. Instead of the thermal pressure one can found the quantum pressure, which corresponds to the well-known quantum Bohm potential. It contains contribution of the spin density along with the spin independent part. The Bloch equation was suggested for classical magnetic moment. It shows evolution of a magnetic moment in an external magnetic field. Hydrodynamical Bloch equation appears for the vector field of the spin density. This equation contains divergence of the spin-current and the torque field caused by an external field. One-particle spin-current consists of two parts: kinematic one, which is the product of the spin density on the velocity field, and the quantum term, which is an analog of the quantum Bohm potential in the Euler equation.

The fact that the one-particle Pauli equation can be presented as the described set of quantum hydrodynamic equations leads to some applications. Spinless part of the hydrodynamic equations looks like classic hydrodynamic equations for a system of charged particle, the classic plasma. Technically there are only two differences between these sets. The first differences is the presence of the quantum Bohm potential instead of the thermal pressure. The second one is an explicit form of the force density appearing as the Lorentz force. However the set of quantum equations obtained from the one-particle Pauli equation contains force caused by an external fields. Whereas in the classical hydrodynamic, corresponding to a many-particle system, we have the force of interaction along with the external force. Internal fields describing interparticle interaction satisfy to the Maxwell equations, where the charge and current densities given by particle evolution.

One can use this similarity of the one-particle quantum hydrodynamics and the hydrodynamic equations of classic plasma \cite{Marklund PRL 07}. For this aim one should account the thermal pressure along with the quantum Bohm potential and assume that he has full force density coupled by the Maxwell equations. Thus, the set of quantum hydrodynamic equations can be used to estimate contribution of spin evolution in properties of quantum plasmas. Nevertheless this manipulation does not necessary. We have mentioned above that direct derivation of the many-particle quantum hydrodynamics was performed. Let us now make a brief description of main ideas used to construct the many-particle hydrodynamic.

To describe a quantum system we need to get quantum average of operators using the wave function. Considering many-particle system we should choose useful operators, and calculate corresponding quantum variables using averaging of the operators. But we have to have a many-particle wave function for physical system under consideration. This is impossible to get the wave function of the many-particle system. However we can construct a suitable many-particle quantum variable via wave function. So we can use the Schrodinger equation to get equation for this function evolution. Knowing this equation we can calculate the many-particle function without knowledge of the wave function. We know that the one-particle Pauli equation can be presented in the hydrodynamic form. It gives us a clue. We can derive a hydrodynamic equation set for many-particle system. And we need to choose suitable many-particle quantum variable. Classical physics gives us a hint. To derive the set of classic hydrodynamic equations we should define the particle concentration. All other functions appears during equations derivation. Therefore we should get operator of the particle concentration. This operator appears as the sum of Dirac delta functions depending on coordinates of particles, each of them is the density of the point-like particle.
Making quantum averaging of this operator with a many-particle wave function we get the quantum particle concentration. In this paper we want to get quantum kinetic equations. So we should get an operator of the distribution function. Thus we have to take classic microscopic distribution function 
replace the coordinate $\textbf{r}_{i}$ and the momentum $\textbf{p}_{i}$ of particles with the corresponding operators $\widehat{\textbf{r}}_{i}=\textbf{r}_{i}$,  $\widehat{\textbf{p}}_{i}=-\imath\hbar\nabla_{i}$. Taking the quantum average of the operator with the many-particle wave function we obtain the quantum microscopic distribution function. Formulas and technical details will be presented in Sec. II. So many-particle quantum hydrodynamics is a representation of the Pauli equation for N-particles in terms of material fields. These ideas were developed in Refs. \cite{MaksimovTMP 1999}, \cite{MaksimovTMP 2001}. Further development of the many-particle quantum hydrodynamics was presented in Refs. \cite{Andreev MPL B 13}-\cite{Andreev Asenjo}.

Another method of derivation of the quantum hydrodynamic equations including energy evolution was suggested in Ref. \cite{Koide PRC 13}. This method uses definition of the momentum density given by Minkowski. The condition of the algebraic positivity of the entropy production was used, at derivation, in the framework of the linear irreversible thermodynamics.

We have discussed the quantum hydrodynamic and kinetic models as the background of our research since we suggest new way of the quantum kinetic equations. Our equations will be applied to study the spin wave in a system of neutral spinning particles. So let us to move to discussion of the spin waves and some results obtained by means of the quantum hydrodynamic.

Spin waves in ferromagnetic and dia- and para-magnetic materials have a great history. Simples example of the spin waves can be found in famous books on basic principles of condensed matter physics \cite{Ashcroft}, \cite{Kittel}.

Magnitude of the spin-spin interaction decreases as $1/r^{3}$. This is faster then decreasing of the Coulomb interaction $1/r$. So, some times, the spin-spin interaction was considered as a short-range interaction \cite{Kittel}. Otherwise comparing the spin-spin interaction with the Lennard-Jones potential $U=c_{12}/r^{12}-c_{6}/r^{6}$, which is the textbook example of the short-range interaction, we can conclude that the spin-spin interaction is an example of the long-range interaction. Assuming this we can use the self-consistent field approximation for the spin-spin interaction. In this approximation results of the many-particle quantum hydrodynamic coincide with the one-particle quantum hydrodynamic up to account of the energy and temperature evolution, which require the many-particle treatment. But many-particle theory allows to get contribution of the quantum correlation, additional to the self-consistent field. An effective radius of the spin-spin interaction smaller then an effective radius of the Coulomb interaction. However in many-particle mixture of electrons and ions one can find the electric-field screening, that the 	
decrease the effective radius of the Coulomb interaction. Thus the Coulomb field of a charge decreases as $e^{-r/r_{d}}/r$, where $r_{d}$ is the screening radius. So we have more similarity between the Coulomb and spin-spin interaction.

Spin waves in degenerate quantum plasmas and contribution of the spin evolution in properties of the magnetized plasma waves have been considered in the last decade \cite{Brodin PRL 08 a}, \cite{Marklund PRL 07}, \cite{Maksimov VestnMSU 2000}, \cite{Andreev VestnMSU 2007}, \cite{Andreev AtPhys 08}, \cite{Asenjo PL A 12}. Quantum hydrodynamics and quantum kinetics were used there. At studying of the spin waves in the quantum plasmas the spin-spin interaction was considered as a long-range interaction. In Ref. \cite{Andreev AtPhys 08} a contribution of the exchange interaction for the Coulomb interaction and the spin-spin interaction in dispersion of the spin and plasma waves was considered. The exchange interaction appears in the quantum hydrodynamic equations as an additional quantum pressure \cite{MaksimovTMP 2001 b}, \cite{Andreev AtPhys 08} (see for instance formulas 2 and 3 of Ref. \cite{Andreev AtPhys 08}). As a consequence we see that the exchange interaction changes the Fermi velocity. In this way it changes dispersion of well-known waves in the magnetized plasmas. The Coulomb exchange interaction in quantum kinetics was recently considered in Ref. \cite{Zamanian 13} in the framework of the Wigner distribution function.

The spin waves in different structures have been studied. This is a huge topic covering many fields in the condensed matter physics, including the dipolar quantum gases. We are not describing all these fields. We restrict ourself with described spin waves in the quantum plasmas.

Quantum vorticity appearing as spin mapping has been considered in terms of quantum hydrodynamics \cite{Andreev Asenjo}, \cite{Asenjo PL A 12}, \cite{Mahajan PRL 11}.

System of the neutrons is an example of the chargeless spinning quantum plasma \cite{Andreev AtPhys 08}, \cite{Mahajan PL A 13}, along with the atoms, which is a more complicated example since they have reach structure of electron levels including the Zeeman splitting.

This paper is organized as follows. In Sect. II we consider basic properties of the model we are developing and using. We present basic definitions and explain how to get kinetic equations from the Schrodinger equation for many-particle system. We also give comparison of results given by our method with the well-known results. We consider charged particles with no spin to get easier comparison. In Sec. III we derive a set of kinetic  equations for neutral spinning particles with the long-range spin-spin interaction. We also introduce the spin distribution function appearing in the kinetic equation for distribution function $f(\textbf{r},\textbf{p},t)$. We present general equations and discuss the self-consistent field approximation allowing to get truncation of the kinetic equations chain. Dispersion of the spin waves in the three dimensional system is considered in Sec. IV. We calculate dispersion dependencies for waves propagating perpendicular to an external magnetic field. We consider waves caused by evolution of the spin distribution function projection on the direction of the external field. In Sec. V we study the spin waves in two dimensional system of neutral particles. We present the explicit form of kinetic equations for the two dimensional samples in the self-consistent field approximation. We obtain two dispersion equations, one appears due to $S_{z}$ evolution, and the second one appears due to $S_{x}$ and $S_{y}$ evolution. Dispersion dependencies are analyzed. Comparison with the three dimensional case is presented. In Sec. VI we
present the brief summary of our results.

\section{II. Model}

\subsection{A. General definitions and properties}

Microscopic density of particles in classical physics is a sum of the Dirac delta functions
\begin{equation}\label{QspinKin concentr def}n(\textbf{r},t)=\sum_{n=1}^{N}\delta(\textbf{r}-\textbf{r}_{n}(t))\end{equation}
each of them presents point like particle \cite{Klimontovich book}-\cite{Kuz'menkov 91}, where $N$ is the total number of particles in the system.
Since a quantum mechanical observable appears as the average of corresponding operator on the wave function of considered system we have to use definition (\ref{QspinKin concentr def}) for obtaining of the quantum concentration of particles. Let us admit that the quantum averaging gives us exact value of variable in considering state. It is not related with the statistical averaging then we do not know details evolution of the system and it is enough to have approximate picture of "average" evolution. Thus definition (\ref{QspinKin concentr def})
was used for construction of the quantum particles
concentration and derivation of the quantum hydrodynamic equations
\cite{MaksimovTMP 1999}, \cite{Andreev PRB 11}, \cite{Andreev IJMP B 12}. Corresponding microscopic distribution function can be
written as
\begin{equation}\label{QspinKin distr func def} f(\textbf{r},\textbf{p},t)=\sum_{n}\delta(\textbf{r}-\textbf{r}_{n}(t))\delta(\textbf{p}-\textbf{p}_{n}(t)),\end{equation}
where $\delta(\textbf{r})$ is the Dirac $\delta$ function.
In classical physics $f(\textbf{r},\textbf{p},t)$ (\ref{QspinKin distr func def}) has been used for derivation
of kinetic equations \cite{Klimontovich book}-\cite{Kuz'menkov 91}. Constructing corresponding quantum
operator and considering its quantum averaging we come to the quantum distribution function.

We have mentioned the quantum averaging for several times.
To be strait we present definition of the quantum mechanical
averaging
$$<L>=\int \psi^{*}\hat{L}\psi dR$$
of a quantum variable $L$ describing by operator $\hat{L}$(see Ref. \cite{Landau Vol 3} or other textbooks on quantum mechanics)

Let us to present the definition of the quantum distribution function
\begin{equation}\label{QspinKin distr func operator intr} \hat{f}=\sum_{n}\delta(\textbf{r}-\widehat{\textbf{r}}_{n})\delta(\textbf{p}-\widehat{\textbf{p}}_{n}),\end{equation}
which is the main definition in this paper,
where $\widehat{\textbf{p}}_{n}$ is the momentum operator for
$n$th particle, and $\textbf{p}$ is the numerical vector function
which arithmetizes the momentum space, as the coordinate
$\textbf{r}$ arithmetizes the coordinate space.

\subsection{B. Kinetic equation for charged spinless particles}

Considering long-range interactions the classic and quantum kinetic method has been
developed for charged particles. Therefore in this section we present our method derivation
of kinetic equation considering system of charged particles. It will give connection of our
results for neutral spinning particles with well-known results of kinetic theory.

The equation of quantum kinetics is derived from the
non-stationary Schrodinger equation for system of N particles:
$$\imath\hbar\partial_{t}\psi(R,t)=\Biggl(\sum_{n}\biggl(\frac{1}{2m_{n}}\widehat{\textbf{p}}_{n}^{2}+e_{n}\varphi_{n,ext}\biggr)$$
\begin{equation}\label{QspinKin Hamiltonian spinless}+\frac{1}{2}\sum_{n,k\neq
k}e_{n}e_{k}G_{nk}\Biggr)\psi(R,t).\end{equation} The following
designations are used in the equation (\ref{QspinKin Hamiltonian spinless}):
$p_{n}^{\alpha}=-\imath\hbar\partial_{n}^{\alpha}$,
$\varphi_{n,ext}$ is the vector potential of an
external electromagnetic field,
$\partial_{n}^{\alpha}=\nabla_{n}^{\alpha}$ is the derivative on
the space variables of $n$th particle, and $G_{nk}=1/r_{nk}$  is
the Green functions of the Coulomb interaction, $\psi(R,t)$ is the
psi function of N particle system,
$R=(\textbf{r}_{1},...,\textbf{r}_{N})$, $e_{n}$, $m_{n}$ are the
charge and the mass of particle, $\hbar$ is the Planck constant
and $c$ is the speed of light.

Using the operator of distribution function (\ref{QspinKin distr func operator intr})
we can obtain the quantum distribution function taking quantum mechanical average of the operator. So we find
$$f(\textbf{r}, \textbf{p},t)=\frac{1}{2}\int \psi^{*}(R,t)\sum_{n}\biggl(\delta(\textbf{r}-\textbf{r}_{n})\delta(\textbf{p}-\widehat{\textbf{p}}_{n})$$
\begin{equation}\label{QspinKin def distribution function non sym}+\delta(\textbf{p}-\widehat{\textbf{p}}_{n})\delta(\textbf{r}-\textbf{r}_{n})\biggr)\psi(R,t)dR,\end{equation}
where $dR=\prod_{n=1}^{N}d\textbf{r}_{n}$. This definition symmetric relatively to operators $\delta(\textbf{r}-\textbf{r}_{n})$ and $\delta(\textbf{p}-\widehat{\textbf{p}}_{n})$, but it is not fully symmetric. To get the fully symmetric definition of the distribution function we need to add complex conjugated quantity. Thus we have following definition
$$f(\textbf{r}, \textbf{p},t)=\frac{1}{4}\int \Biggl(\psi^{*}(R,t)\sum_{n}\biggl(\delta(\textbf{r}-\textbf{r}_{n})\delta(\textbf{p}-\widehat{\textbf{p}}_{n})$$
\begin{equation}\label{QspinKin def distribution function}+\delta(\textbf{p}-\widehat{\textbf{p}}_{n})\delta(\textbf{r}-\textbf{r}_{n})\biggr)\psi(R,t)+h.c.\Biggr)dR,\end{equation}
where h.c. stands for the complex conjugation.

Integrating of the distribution function over the momentum we have the particle concentration
\begin{equation}\label{QspinKin concentration}n(\textbf{r},t)=\int f(\textbf{r}, \textbf{p},t) d\textbf{p},\end{equation}
which coincides with the definition used in the many-particle quantum hydrodynamics \cite{MaksimovTMP 1999}, \cite{Andreev PRB 11} and has following form
\begin{equation}\label{QspinKin def density}n(\textbf{r},t)=\int dR\sum_{n}\delta(\textbf{r}-\textbf{r}_{n})\psi^{*}(R,t)\psi(R,t).\end{equation}

The distribution function (\ref{QspinKin def
distribution function}) satisfies to the following quantum kinetic equation
\begin{equation}\label{QspinKin kinetic equation charge}\partial_{t}f+\frac{1}{m}\textbf{p}\partial_{\textbf{r}}f+e\frac{\imath}{\hbar}\varphi(\textbf{r},t)\sin(\overleftarrow{\nabla}_{\textbf{r}}\nabla_{\textbf{p}})f=0,\end{equation}
where
\begin{equation}\label{QspinKin sin defenition} \sin(\overleftarrow{\nabla}_{\textbf{r}}\nabla_{\textbf{p}})=\sum_{l=1}^{\infty}\frac{(\imath\hbar)^{2l+1}}{(2l+1)!}(\overleftarrow{\nabla}_{\textbf{r}}\nabla_{\textbf{p}})^{2l+1}.\end{equation}
In equation (\ref{QspinKin sin defenition}) we have
used designation
$\varphi(\textbf{r},t)\overleftarrow{\nabla}_{\textbf{r}}$, it
means that the gradient operator$\nabla$ on spatial variables acts
in the left-hand side, instead of usual acting of operators on
function standing in the right-hand side. We do not write the
Plank constant $\hbar$ in the argument of $\sin$ to make this
notation more handy.

At studying of spin waves we will use the first term in the series (\ref{QspinKin sin defenition}) only.
\begin{equation}\label{QspinKin sin def approx} \sin(\overleftarrow{\nabla}_{\textbf{r}}\nabla_{\textbf{p}})\simeq \imath\hbar\overleftarrow{\nabla}_{\textbf{r}}\nabla_{\textbf{p}}.\end{equation}
That corresponds to the quasi classical approximation.
To make this paper easier to read and to be closer to application to wave dispersion we will use approximation (\ref{QspinKin sin def approx}) in all general equations below.

As kinetic methods have been mostly used for studying of charged particles we presented derivation of the quantum kinetic equations for charged spinless particles. We did it to show that our method gives expected results for plasmas.

Equation (\ref{QspinKin kinetic equation charge}) in the quasi classical approximation looks as
\begin{equation}\label{QspinKin kinetic equation charge appr}\partial_{t}f+\frac{1}{m}\textbf{p}\partial_{\textbf{r}}f+e\textbf{E} \nabla_{\textbf{p}}f=0,\end{equation}
where $\textbf{E}=-\nabla_{\textbf{r}}\varphi$. We see that equation (\ref{QspinKin kinetic equation charge appr}) coincides with the Vlasov equation. We can conclude that the distribution function used in this section gives reasonable results. Next, we can use this definition (\ref{QspinKin def distribution function}) to develop a kinetic theory of spinning particles.

\section{III. Kinetic evolution of chargeless spinning particles}

In section II we present the Schrodinger equation contained
the Coulomb interaction only. Now we are going to consider kinetics of chargeless
spinning particles. We focus our attention on spin-1/2 fermions. We suppose that particles interact by mean of the spin-spin interactions.
In this way a particle having spin and corresponding magnetic moment creates
the magnetic field acting on surrounding particles.
For this model we have the following Hamiltonian

$$\hat{H}=\sum_{n}\biggl(\frac{1}{2m_{n}}\widehat{\textbf{p}}^{2}_{n}-\gamma_{n}\widehat{\sigma}^{\alpha}_{n}B^{\alpha}_{n(ext)}\biggr)$$
\begin{equation}\label{QspinKin Ham spinning part}-\frac{1}{2}\sum_{k,n\neq p}\gamma_{k}\gamma_{n}G^{\alpha\beta}_{kn}\widehat{\sigma}^{\alpha}_{k}\widehat{\sigma}^{\beta}_{n},\end{equation}
where $\hat{p}_{n}^{\alpha}=-\imath\hbar\partial_{n}^{\alpha}$,
the Green
function of the spin-spin interaction has the following form
$G^{\alpha\beta}_{pn}=4\pi\delta^{\alpha\beta}\delta(\textbf{r}_{pn})+\partial^{\alpha}_{p}\partial^{\beta}_{p}(1/r_{pn})$$=(\partial^{\alpha}_{p}\partial^{\beta}_{p}-\delta^{\alpha\beta}\triangle_{p})(1/r_{pn})$,
$\gamma_{p}$ is the gyromagnetic ratio. $B^{\alpha}_{
(ext)}(\textbf{r}_{p},t)$ is the external magnetic field,
$\widehat{\sigma}^{\alpha}_{p}$ are the Pauli matrix, the commutation
relations for them is
$$[\widehat{\sigma}^{\alpha}_{p},\widehat{\sigma}^{\beta}_{n}]=2\imath\delta_{pn}\varepsilon^{\alpha\beta\gamma}\widehat{\sigma}^{\gamma}_{p}.$$
We use it for derivation of the fundamental kinetic equations.

To obtain quantum kinetic equation for the distribution function (\ref{QspinKin def distribution function}) we differentiate the explicit form of the distribution function (\ref{QspinKin def distribution function}) with respect to time and use the Schrodinger equation with the Hamiltonian (\ref{QspinKin Ham spinning part}) for the time derivatives of the many-particle wave function. After some straightforward calculations we obtain the kinetic equation presented below.

For simplicity we present the kinetic equation for spinning neutral particles in the absence of the inter-particle interaction
\begin{equation}\label{QspinKin kinetic equation gen with spin semi classic limit}\partial_{t}f+\frac{\textbf{p}}{m}\partial_{\textbf{r}}f
+\partial_{\alpha} B^{\beta}_{ext}(\textbf{r},t)\partial_{\textbf{p}\alpha} S^{\beta}(\textbf{r}, \textbf{p},t)=0,\end{equation}
where we have new quantity in these equations, we can call it spin-distribution function, its explicit form is
$$S^{\alpha}(\textbf{r}, \textbf{p},t)=\frac{1}{4}\int \Biggl(\psi^{*}(R,t)\sum_{n}\biggl(\delta(\textbf{r}-\textbf{r}_{n})\delta(\textbf{p}-\widehat{\textbf{p}}_{n})$$
\begin{equation}\label{QspinKin def spin distribution function}+\delta(\textbf{p}-\widehat{\textbf{p}}_{n})\delta(\textbf{r}-\textbf{r}_{n})\biggr)\sigma^{\alpha}_{n}\psi(R,t)+h.c.\Biggr)dR,\end{equation}
where h.c. stands for the Hermitian conjugation.
$S^{\alpha}(\textbf{r}, \textbf{p},t)$ is kinetic analog of the spin density, which arises in the quantum hydrodynamics \cite{MaksimovTMP 2001} and has form
\begin{equation}\label{QspinKin def spin density}S^{\alpha}(\textbf{r},t)=\int dR\sum_{n}\delta(\textbf{r}-\textbf{r}_{n})\psi^{*}(R,t)\widehat{\sigma}^{\alpha}_{n}\psi(R,t),\end{equation}
we have used same letter for designation of the spin density and the spin-distribution function, but they differ by set of arguments.

Integrating the spin distribution function over the momentum we get the spin density appearing in the quantum hydrodynamic equations \cite{Andreev IJMP B 12}, \cite{Andreev spin current}, \cite{Andreev Asenjo}
\begin{equation}\label{QspinKin def spin density connection of S and S} S^{\alpha}(\textbf{r}, t)=\int S^{\alpha}(\textbf{r}, \textbf{p},t)d\textbf{p}.\end{equation}
Magnetization $M^{\alpha}(\textbf{r},t)$ usually used in the
quantum hydrodynamics \cite{MaksimovTMP 2001}, \cite{Andreev IJMP B 12}, and \cite{Andreev spin current}. Magnetization $M^{\alpha}(\textbf{r},t)$ has simple
connection with the spin density $M^{\alpha}(\textbf{r},t)=\gamma
S^{\alpha}(\textbf{r},t)$, where $\gamma$ is the gyromagnetic
ratio for considering species of particles.

To get complete description of  systems of spinning particles we
have to derive an equation for the spin-distribution function
$S^{\alpha}(\textbf{r}, \textbf{p},t)$.

\subsection{Including of interaction}

Introducing of interaction makes  the kinetic equation more larger. It gives one more term describing the spin-spin interaction.
$$\partial_{t}f+\frac{1}{m}\textbf{p}\partial_{\textbf{r}}f+\gamma (\nabla_{\textbf{r}}^{\beta}B^{\alpha}_{ext})\nabla_{\textbf{p}}^{\beta}S^{\alpha}(\textbf{r}, \textbf{p},t)$$
\begin{equation}\label{QspinKin kinetic equation gen with spin and int}-\gamma^{2}
\int d\textbf{r}' (\nabla_{\textbf{r}}^{\alpha}G^{\beta\gamma}(\textbf{r},\textbf{r}'))\nabla_{\textbf{p}}^{\alpha} S_{2}^{\beta\gamma}(\textbf{r},\textbf{p},\textbf{r}',\textbf{p}',t)=0. \end{equation}
The last term describing interaction contains a two-particle function $S^{\alpha\beta}_{2}$. We will consider how to deal with two-particle functions below.
In kinetic equation for spinning particles appears the spin distribution function. Therefore, for construction of the closed set of equation describing spinning particles we have to find equation evolution of the spin distribution function. For this goal we differentiate spin distribution function with respect to time, after some calculations we find the kinetic equation for spin distribution function evolution
$$\partial_{t}S^{\alpha}(\textbf{r},\textbf{p},t)+\frac{1}{m}\textbf{p}\partial_{\textbf{r}}S^{\alpha}
+\gamma (\nabla^{\beta} B^{\alpha}_{ext}) \nabla_{\textbf{p}}^{\beta} f(\textbf{r}, \textbf{p},t)$$
$$-\gamma^{2}
\int (\nabla_{\textbf{r}}^{\gamma}G^{\alpha\beta}(\textbf{r},\textbf{r}'))  \nabla_{\textbf{p}}^{\gamma} N^{\beta}_{2}(\textbf{r},\textbf{p},\textbf{r}',\textbf{p}',t) d\textbf{r}'d\textbf{p}'$$
$$-\frac{2\gamma}{\hbar}\varepsilon^{\alpha\beta\gamma}\Biggl(S^{\beta}(\textbf{r},\textbf{p},t)B^{\gamma}_{ext}(\textbf{r},t)$$
\begin{equation}\label{QspinKin kinetic equation (spin evol) gen with spin and int with two part F}+\gamma\int G^{\gamma\delta}(\textbf{r},\textbf{r}')S_{2}^{\beta\delta}(\textbf{r},\textbf{p},\textbf{r}',\textbf{p}',t)d\textbf{r}'d\textbf{p}'\Biggr)=0.\end{equation}
Equations (\ref{QspinKin kinetic equation gen with spin and int}) and
(\ref{QspinKin kinetic equation (spin evol) gen with spin and int with two part F}) are the first two equations of the chain of quantum kinetic equations.
To get a closed set of equation we have to truncate this chain.
To make truncation we need to consider the two-particle distribution functions appeared in equations
(\ref{QspinKin kinetic equation gen with spin and int}) and (\ref{QspinKin kinetic equation (spin evol) gen with spin and int with two part F}).
Let us present the explicit definitions of the two-particle functions.
$$S_{2}^{\alpha\beta}(\textbf{r},\textbf{p},\textbf{r}',\textbf{p}',t)$$
$$=\frac{1}{4}\int dR\Biggl(\psi^{*}\sum_{n,k\neq n}\delta(\textbf{r}'-\textbf{r}_{k})\delta(\textbf{p}'-\widehat{\textbf{p}}_{k})\times$$
\begin{equation}\label{QspinKin spin distrib func two part}\times\biggl(\delta(\textbf{r}-\textbf{r}_{n})\delta(\textbf{p}-\widehat{\textbf{p}}_{n})+\delta(\textbf{p}-\widehat{\textbf{p}}_{n})\delta(\textbf{r}-\textbf{r}_{n})\biggr)\sigma_{n}^{\alpha}\sigma_{k}^{\beta}\psi+h.c.\Biggr) \end{equation}
is the two-particle spin-spin distribution function. In the absence of correlations, i.e. in the self-consistent field approximation, this function splits
on the product of two spin distribution functions;
and
$$N_{2}^{\alpha}(\textbf{r},\textbf{p},\textbf{r}',\textbf{p}',t)$$
$$=\frac{1}{4}\int dR\Biggl(\psi^{*}\sum_{n,k\neq n}\delta(\textbf{r}'-\textbf{r}_{k})\delta(\textbf{p}'-\widehat{\textbf{D}}_{k})\times$$
\begin{equation}\label{QspinKin two part distr func N}\times\biggl(\delta(\textbf{r}-\textbf{r}_{n})\delta(\textbf{p}-\widehat{\textbf{D}}_{n})+\delta(\textbf{p}-\widehat{\textbf{D}}_{n})\delta(\textbf{r}-\textbf{r}_{n})\biggr)\sigma_{k}^{\alpha}\psi+h.c.\Biggr)\end{equation}
is the two-particle position-spin distribution function splitting, in the self-consistent field approximation, on the product of
the distribution function $f(\textbf{r},\textbf{p},t)$ and the spin distribution function $\textbf{S}(\textbf{r},\textbf{p},t)$.
Two-particle distribution functions contain a contribution of the exchange interaction, which gives no trace in the self-consistent field
approximation. So one can notice that the self-consistent field approximation is related to the Hartree approximation
in the quantum mechanics. If we want to consider an analog of the Hartree–Fock approximation we need to have more
general formulas for two-particle distribution functions.

Only interaction we consider in the paper is the spin-spin interaction, which is an example of the long-range interaction. The self-consistent field approximation
is very suitable approximation for the long-range interaction. This approximation means that
we should present the two-particle function as the product of corresponding one-particle functions. Obtained here quantum kinetic equations
(\ref{QspinKin kinetic equation gen with spin and int}) and (\ref{QspinKin kinetic equation (spin evol) gen with spin and int with two part F})
contain two-particle functions (\ref{QspinKin spin distrib func two part}) and (\ref{QspinKin two part distr func N}).
In the self-consistent field approximation they can be written as
\begin{equation}\label{QspinKin def distribution spin two part function self consist field appr}S_{2}^{\alpha\beta}(\textbf{r},\textbf{p},\textbf{r}',\textbf{p}',t)=S^{\alpha}(\textbf{r},\textbf{p},t)S^{\beta}(\textbf{r}',\textbf{p}',t),\end{equation}
and
\begin{equation}\label{QspinKin def distribution N2 self consist field}N_{2}^{\alpha}(\textbf{r},\textbf{p},\textbf{r}',\textbf{p}',t)=f(\textbf{r},\textbf{p},t) S^{\alpha}(\textbf{r}',\textbf{p}',t).\end{equation}

In the result we have next set of equations
\begin{equation}\label{QspinKin kinetic equation gen  classic limit with E and B} \partial_{t}f+\frac{\textbf{p}}{m}\partial_{\textbf{r}}f+\partial_{\alpha} B^{\beta}(\textbf{r},t)\partial_{\textbf{p}\alpha} S^{\beta}(\textbf{r}, \textbf{p},t)=0,\end{equation}
and
\begin{equation}\label{QspinKin kinetic equation gen for spin classic limit with E and B} \partial_{t}S^{\alpha}+\frac{\textbf{p}}{m}\partial_{\textbf{r}}S^{\alpha}+\partial_{\beta} B^{\alpha}(\textbf{r},t)\partial_{\textbf{p}\beta} f-\frac{2\gamma}{\hbar}\varepsilon^{\alpha\beta\gamma}S^{\beta}B^{\gamma}=0,\end{equation}
where magnetic field $\textbf{B}$ is the full magnetic field, which is the sum of the external $\textbf{B}_{ext}$ and the internal
\begin{equation}\label{QspinKin def of B} B^{\alpha}_{int}=\gamma\int  G^{\alpha\beta}(\textbf{r},\textbf{r}') S^{\beta}(\textbf{r}',\textbf{p}',t) d\textbf{r}' d\textbf{p}' \end{equation} is the magnetic field created by magnetic moments (spins).


Due to the fact that we have considered spin-spin interaction and we have not included spin-current and current-current interaction we have magnetic field satisfying to the following equations
$$\nabla\textbf{B}=0,$$
and
\begin{equation}\label{QspinKin electro stat Max in spin chapter} \begin{array}{ccc}\nabla\times \textbf{B}=4\pi\nabla\times\int \textbf{S}(\textbf{r},\textbf{p},t)d\textbf{p}.&   \end{array}\end{equation}

Set of equations (\ref{QspinKin kinetic equation gen  classic limit with E and B})-(\ref{QspinKin electro stat Max in spin chapter}) is a closed
set of quantum kinetic equations for spinning particles. The third term in equation (\ref{QspinKin kinetic equation gen  classic limit with E and B})
describes evolution of the distribution function under influence of magnetic field $\textbf{B}$ acting on the magnetic moments (spins).

The last term in equation (\ref{QspinKin kinetic equation gen  classic limit with E and B}) contains
contribution of the third and fourth terms of equation (\ref{QspinKin kinetic equation gen with spin and int}).
Analogously, the third term in equation (\ref{QspinKin kinetic equation gen for spin classic limit with E and B})
presents the third and fourth terms of equation (\ref{QspinKin kinetic equation (spin evol) gen with spin and int with two part F}).
And the last term of equation (\ref{QspinKin kinetic equation gen for spin classic limit with E and B}) includes contribution
of the last group of terms in equation
(\ref{QspinKin kinetic equation (spin evol) gen with spin and int with two part F}).

Integration of equation (\ref{QspinKin kinetic equation gen  classic limit with E and B}) over the momentum gives the continuity equation \cite{Andreev spin current}, \cite{Klimontovich book}. The first (second) term gives $\partial_{t}n$ ($\nabla\textbf{j}=\nabla(n\textbf{v})$), where $n$ is the particle concentration, $\textbf{v}$ is the velocity field, and $\textbf{j}$ is the momentum density. The last tern of equation (\ref{QspinKin kinetic equation gen  classic limit with E and B}) gives no contribution in the continuity equation. Multiplying equation (\ref{QspinKin kinetic equation gen  classic limit with E and B}) on the momentum and integration it over the momentum we come to the Euler equation or, in other words, the momentum balance equation \cite{Andreev spin current}, \cite{Klimontovich book}. The first (second) term of equation (\ref{QspinKin kinetic equation gen  classic limit with E and B}) gives $\partial_{t}\textbf{j}$ ($\partial^{\beta}\Pi^{\alpha\beta}$), where $\Pi^{\alpha\beta}=nv^{\alpha}v^{\beta}+p^{\alpha\beta}+T^{\alpha\beta}$ is the momentum flux, $p^{\alpha\beta}$ is the tensor of thermal pressure, $T^{\alpha\beta}$ is the quantum Bohm potential \cite{Andreev Asenjo}. The last term in equation (\ref{QspinKin kinetic equation gen  classic limit with E and B}) gives the force density $\textbf{F}=M^{\beta}\nabla B^{\beta}$, which is well known in the quantum hydrodynamics \cite{MaksimovTMP 2001}, \cite{Marklund PRL 07}, \cite{Andreev Asenjo}.

Integration of equation (\ref{QspinKin kinetic equation gen for spin classic limit with E and B}) over the momentum leads to the Bloch equation (spin or magnetic moment evolution equation) \cite{MaksimovTMP 2001}, \cite{Andreev spin current}. The first (second) term in equation (\ref{QspinKin kinetic equation gen for spin classic limit with E and B}) leads to $\partial_{t}\textbf{M}$ ($\partial_{\beta}J^{\alpha\beta}$), where $\textbf{M}$ is the density of magnetic moments, and $J^{\alpha\beta}$ is the spin-current \cite{Andreev spin current}. The third term disappears at integrating over the momentum. It can reveal in the spin-current evolution equation, which appears at multiplying of equation (\ref{QspinKin kinetic equation gen for spin classic limit with E and B}) on the momentum and integration of obtained tensor equation over the momentum. In this way it appears at kinetic treatment. It was shown in Ref. \cite{Andreev spin current} that the spin-current evolution equation can be derived directly from the Schrodinger equation as a part of the quantum hydrodynamic equations set. Thus we can conclude that the third term gives some additional information in comparison with the standard quantum hydrodynamical model of spinning particles \cite{MaksimovTMP 2001}, \cite{Marklund PRL 07}, \cite{Andreev IJMP B 12}. The last term in equation (\ref{QspinKin kinetic equation gen for spin classic limit with E and B}) is an kinetic analog of the torque or moment of force acting on the magnetic moment in the Bloch equation. Comparing with the quantum hydrodynamic we can point out that in hydrodynamic we have the  cross product of the spin-density $\textbf{S}(\textbf{r},t)$ and the magnetic field $\textbf{B}(\textbf{r},t)$. While in the quantum kinetic we have the cross product of the spin distribution function $\textbf{S}(\textbf{r},\textbf{p},t)$ and the magnetic field $\textbf{B}(\textbf{r},t)$.

\section{IV. Dispersion of spin waves in three dimensional systems}

We consider system of neutral spinning particles in an external uniform magnetic field. We study evolution of small perturbations of an equilibrium state described by a distribution function $f_{0}(\textbf{p})$, a spin distribution function $\textbf{S}_{0}(\textbf{p})$, and external magnetic field $\textbf{B}_{0}$. The equilibrium spin distribution function is a vector parallel to the external magnetic field $\textbf{S}_{0}(\textbf{p})\parallel\textbf{B}_{0}$. We assume that perturbation of the distribution function $\delta f$, the spin distribution function $\delta \textbf{S}$, and the magnetic field $\delta\textbf{B}$ are monochromatic waves and they are proportional to $\exp\biggl(-\imath\omega t+ \imath \textbf{k} \textbf{r}\biggr)$. We suppose that the external field is parallel to $z$ axes $\textbf{B}_{0}\parallel \textbf{e}_{z}$.

\subsection{A. Calculation of three dimensional spectrum}

Linearized set of the kinetic (\ref{QspinKin kinetic equation gen  classic limit with E and B}), (\ref{QspinKin kinetic equation gen for spin classic limit with E and B}) and Maxwell (\ref{QspinKin electro stat Max in spin chapter}) equations are
\begin{equation}\label{QspinKin kin eq f lin 3D}-\imath\omega\delta f
+\imath \frac{\textbf{p}\textbf{k}}{m}\delta f+\imath\gamma \textbf{k}\delta B_{z}\nabla_{\textbf{p}}S_{0}(\textbf{p})=0,\end{equation}
$$-\imath\omega\delta \textbf{S}+\imath \frac{\textbf{p}\textbf{k}}{m} \delta \textbf{S}+\imath\gamma (\textbf{k}\nabla_{\textbf{p}})f_{0}(\textbf{p})\delta \textbf{B}$$
\begin{equation}\label{QspinKin kin eq S lin 3D}+\frac{2\gamma}{\hbar}\biggl(\textbf{B}_{0}\times\delta \textbf{S}-\textbf{S}_{0}\times \delta \textbf{B}\biggr)=0,\end{equation}
\begin{equation}\label{QspinKin Max lin div}\textbf{k}\delta \textbf{B}=0,\end{equation}
and
\begin{equation}\label{QspinKin Max lin curl}\textbf{k}\times\delta \textbf{B}=4\pi\gamma \textbf{k}\times\int\delta \textbf{S}(\textbf{r},\textbf{p},t)d\textbf{p}.\end{equation}

Evolution of the distribution function $f$ does not lead to changes of the spin distribution function $\textbf{S}$ or creation of electromagnetic field. Evolution of the spin distribution function $\textbf{S}$ gives contribution in evolution of the distribution function $f$, but we do not interested in it since it gives no influence in the spin evolution. Let us pay all attention to spin evolution described by the spin distribution function $\textbf{S}$.

Let us put down equations for projections of the spin distribution function $\textbf{S}$
\begin{equation}\label{QspinKin Sx lin 3D}-\imath \varpi\delta S_{x}-\Omega_{\gamma}\delta S_{y}=-\imath\gamma(\textbf{k}\nabla_{\textbf{p}})f_{0}\delta B_{x}-\frac{2\gamma}{\hbar}S_{0}\delta B_{y},\end{equation}
\begin{equation}\label{QspinKin Sy lin 3D}\Omega_{\gamma}\delta S_{x}-\imath \varpi\delta S_{y}=-\imath\gamma(\textbf{k}\nabla_{\textbf{p}})f_{0}\delta B_{y}+\frac{2\gamma}{\hbar}S_{0}\delta B_{x},\end{equation}
\begin{equation}\label{QspinKin Sz lin 3D}-\imath \varpi\delta S_{z}=-\imath\gamma(\textbf{k}\nabla_{\textbf{p}})f_{0}\delta B_{z},\end{equation}
where $\varpi=\omega-\textbf{p}\textbf{k}/m$, and $\Omega_{\gamma}=\frac{2\gamma}{\hbar}B_{0}$ is the cyclotron frequency. Having explicit form of equations (\ref{QspinKin Sx lin 3D})-(\ref{QspinKin Sz lin 3D}), it will be easier to follow how we obtain dispersion equations.

\subsection{B. Perpendicular propagation}

Solving equations (\ref{QspinKin kin eq f lin 3D})-(\ref{QspinKin Max lin curl}) assuming that waves propagate perpendicular to an external magnetic field, and putting formulas for $S_{x}$, $S_{y}$, $S_{z}$ in the Maxwell equations (\ref{QspinKin Max lin div}), (\ref{QspinKin Max lin curl}), we come to the two following dispersion equations
\begin{equation}\label{QspinKin Disp eq Perp prop z} 1-4\pi\gamma^{2}\int\frac{\textbf{k}\nabla_{\textbf{p}}f_{0}}{\omega-\textbf{p}\textbf{k}/m}d\textbf{p}=0,\end{equation}
and
$$1-4\pi\gamma^{2}\int\frac{(\omega-\textbf{p}\textbf{k}/m)\textbf{k}\nabla_{\textbf{p}}f_{0}(\textbf{p})}{(\omega-\textbf{p}\textbf{k}/m)^{2}-\Omega_{\gamma}^{2}}d\textbf{p}$$
\begin{equation}\label{QspinKin Disp eq Perp prop xy} +8\pi\frac{\gamma^{2}}{\hbar}\Omega_{\gamma}\int\frac{S_{0}(\textbf{p})}{(\omega-\textbf{p}\textbf{k}/m)^{2}-\Omega_{\gamma}^{2}}d\textbf{p}=0.\end{equation}
We assumed that the external magnetic field is directed parallel to Oz axes. We choose direction of wave propagation parallel to Ox axes. Equation (\ref{QspinKin Disp eq Perp prop z}) appears due to evolution of $\delta B_{z}$ and $\delta S_{z}$. Evolution of $\delta S_{x}$, $\delta S_{y}$, and $\delta B_{y}$ ($\delta B_{x}=0$) leads to equation (\ref{QspinKin Disp eq Perp prop xy}).

To get an explicit form of dispersion equations (\ref{QspinKin Disp eq Perp prop z}) and (\ref{QspinKin Disp eq Perp prop xy}) we should use an explicit form of the equilibrium distribution functions $f_{0}(\textbf{p})$ and $\textbf{S}_{0}(\textbf{p})=S(\textbf{p})_{0}\textbf{e}_{z}$.
\begin{equation}\label{QspinKin maxwell distrib}f_{0}(\textbf{p})=\frac{n_{0}}{(\sqrt{2\pi mT})^{3}}\exp\biggl(-\frac{\textbf{p}^{2}}{2mT}\biggr),\end{equation}
and
\begin{equation}\label{QspinKin eq spin distr func}S_{0}(\textbf{p})=\sigma f_{0}(\textbf{p}),\end{equation}
where $n_{0}$ is the equilibrium particle concentration, $T$ is the temperature, and formula (\ref{QspinKin maxwell distrib}) presents the Maxwell distribution function.

Equation (\ref{QspinKin Disp eq Perp prop z}) leads to following dispersion dependence
\begin{equation}\label{QspinKin Disp eq Perp prop z via Z}1-4\pi\gamma^{2}\frac{n_{0}}{T}\biggl(1+\alpha Z(\alpha)\biggr)=0,\end{equation}
where
\begin{equation}\label{QspinKin}\alpha=\frac{\omega}{kv_{T}},\end{equation}
\begin{equation}\label{QspinKin}v_{T}=\sqrt{\frac{2T}{m}},\end{equation}
and
$$Z(\alpha)=\frac{1}{\sqrt{\pi}}\int_{-\infty}^{+\infty}\frac{\exp(-\xi^{2})}{\xi-\alpha}d\xi$$
\begin{equation}\label{QspinKin Z func definition}=\frac{1}{\sqrt{\pi}}\biggl[ P\int_{-\infty}^{+\infty}\frac{\exp(-\xi^{2})}{\xi-\alpha}d\xi\biggr]+\imath\sqrt{\pi}\exp(-\alpha^{2}),\end{equation}
where the symbol $P$ denotes the principle part of the integral.
Let us present assumptions of this formula.
At $\alpha\gg 1$ we have
\begin{equation}\label{QspinKin Z big alpha} Z(\alpha)\simeq-\frac{1}{\alpha}\biggl(1+\frac{1}{2\alpha^{2}}+\frac{3}{4\alpha^{4}}+...\biggr)+\imath\sqrt{\pi}\exp(-\alpha^{2}),\end{equation} In opposite limit we get
\begin{equation}\label{QspinKin Z small alpha} Z(\alpha)=-2\alpha\biggl(1-\frac{2\alpha^{2}}{3}+...\biggr)+\imath\sqrt{\pi}\exp(-\alpha^{2}).\end{equation}
Using these general formulas we can consider dispersion equation (\ref{QspinKin Disp eq Perp prop z via Z}) and other dispersion equations we obtain below.

At $\alpha\gg 1$ equation (\ref{QspinKin Disp eq Perp prop z via Z}) gives negative solutions only. In leading order on $\alpha$ we find
\begin{equation}\label{QspinKin omega--lambda}\omega^{2}=-\frac{4\pi\gamma^{2}n_{0}k^{2}}{m}.\end{equation}
We see that evolution of $\delta S_{z}$ leads to dispersion equation giving no solution. Comparing with the quantum hydrodynamic description we can admit that such solution does not appear at consideration of usually used set of the continuity, Euler, spin evolution equations. However we can get this solution taking into account the spin-current evolution equation \cite{Andreev spin current}.

Considering equation (\ref{QspinKin Disp eq Perp prop z via Z}) at $\alpha\ll 1$ we find a small frequency solution
\begin{equation}\label{QspinKin Sz 3D small alpha}\omega^{2}_{Re}=\frac{1}{2}(kv_{T})^{2}\biggl(1-\frac{T}{4\pi\gamma^{2}n_{0}}\biggr)\end{equation}
existing under condition $T<4\pi\gamma^{2}n_{0}$ and $T\approx 4\pi\gamma^{2}n_{0}$ \emph{and} revealing an instability
\begin{equation}\label{QspinKin}\nu=-\frac{\sqrt{2\pi}}{4}\frac{kv_{T}}{\biggl(1-\frac{T}{4\pi\gamma^{2}n_{0}}\biggr)},\end{equation}
where $\omega_{Re}$ is given by formula (\ref{QspinKin Sz 3D small alpha}) and $\omega=\omega_{Re}-\imath\nu$.

Calculating integrals in equation (\ref{QspinKin Disp eq Perp prop xy}) we come to
$$1-4\pi\gamma^{2} \frac{n_{0}}{T}\biggl(1+\frac{1}{2}\alpha_{-}Z(\alpha_{-})+\frac{1}{2}\alpha_{+}Z(\alpha_{+})\biggr)$$
\begin{equation}\label{QspinKin Sx Sy disp general 3D perp} +\frac{4\pi\gamma^{2}}{\hbar}\frac{\sigma n_{0}}{kv_{T}}(Z(\alpha_{+})-Z(\alpha_{-}))=0,\end{equation}
where $\alpha_{-}=(\omega-\Omega_{\gamma})/kv_{T}$, and $\alpha_{+}=(\omega+\Omega_{\gamma})/kv_{T}$.

At $\alpha_{+}\gg 1$ and $\alpha_{-}\gg 1$
we obtain
\begin{equation}\label{QspinKin perp pr 01 01} \omega^{2}=\Omega_{\gamma}^{2}-\frac{8\pi\sigma n_{0}\gamma^{2}}{\hbar}\Omega_{\gamma},\end{equation}
for real part of the frequency.
Solution (\ref{QspinKin perp pr 01 01}) is obtained in the leading order on $\alpha_{+}$ and $\alpha_{-}$, so we have used the first term in the big brackets in formula (\ref{QspinKin Z big alpha}).

Imaginary part of the frequency corresponding to solution (\ref{QspinKin perp pr 01 01}) is
$$\nu=\frac{\Omega_{\gamma}}{kv_{T}\omega_{R}}\biggl(\frac{4\pi\sigma n_{0}\gamma^{2}}{\hbar}\biggr)^{2}\times$$
$$\times\biggl[ \biggl(1+\frac{(\omega_{0}-\Omega_{\gamma})\hbar}{2\sigma T}\biggr) \exp\biggl(-\frac{(\omega_{0}-\Omega_{\gamma})^{2}}{k^{2}v_{T}^{2}}\biggr)$$
\begin{equation}\label{QspinKin damping XY perp 3D}+\biggl(1+\frac{(\omega_{0}+\Omega_{\gamma})\hbar}{2\sigma T}\biggr)\exp\biggl(-\frac{(\omega_{0}+\Omega_{\gamma})^{2}}{k^{2}v_{T}^{2}}\biggr)\biggr].\end{equation}
It reveals Landau damping of spin waves.

At $\alpha_{+}\gg 1$ and $\alpha_{-}\ll 1$, in leading order on $\alpha_{+}$ and $\alpha_{-}$, equation appears in the following form
\begin{equation}\label{QspinKin}2D\alpha_{+}\alpha_{-}+(1-F)\alpha_{+}-D=0,\end{equation}
where
\begin{equation}F=2\pi\gamma^{2}\frac{n_{0}}{T},\end{equation}
and
\begin{equation}D=\frac{4\pi\gamma^{2}}{\hbar}\frac{\sigma n_{0}}{kv_{T}},\end{equation}
and we get the following solution
$$\omega=\frac{1}{4D}\Biggl(\biggl[k^{2}v_{T}^{2}(1-F)^{2}+16D^{2}\Omega_{\gamma}^{2}-kv_{T}(1-F)$$
\begin{equation}\label{QspinKin perp pr 02 01}-8D kv_{T}[(1-F)\Omega_{\gamma}-Dkv_{T}]\biggr]^{1/2}-kv_{T}(1-F)\Biggr).\end{equation}
In small wave vector limit $k\rightarrow 0$ formula (\ref{QspinKin perp pr 02 01})
gives the cyclotron frequency
\begin{equation}\label{QspinKin}\omega=\mid\Omega_{\gamma}\mid.\end{equation}
Imaginary part of frequency corresponding to solution (\ref{QspinKin perp pr 02 01}) appears to be
\begin{equation}\label{QspinKin}\nu=-\frac{\sqrt{\pi}}{2}kv_{T}\biggl(\sigma+\frac{\hbar(\omega_{R}-\Omega_{\gamma})}{mv_{T}^{2}}\biggr),\end{equation}
where $\sigma+\hbar(\omega_{R}-\Omega_{\gamma})/(2T)<< 1$ so $\nu\ll kv_{T}\ll\omega_{R}$ and we have a slow instability. This formula gives an instability $\nu<0$. However, the collisional damping, see Ref. \cite{Klimontovich book}, exceeds obtained instability. So we actually have a stable spin wave solution with decreased damping due to spin effect.

Comparing collisional damping and an instability related to the Landau damping on his nature, we can compare these two mechanisms of damping. Considering interparticle interaction we pick out two opposite limit of interaction, they are the long-range and the short-range interactions. Sometimes the short-range interaction can be considered as collision, and it leads to damping of waves. This fact was well-known for a long time, however It came as a surprise that long-range interaction leads to a damping either (Landau damping caused by the Coulomb interaction found in 1946). Thus we conclude that all type of interaction can lead to damping of processes in many-particle systems.

\subsection{C. Parallel propagation}

Considering propagation of waves parallel to external magnetic field we have $\textbf{k}=k \textbf{e}_{z}$. Consequently, we have $\delta B_{z}=0$ from equation (\ref{QspinKin Max lin div}). Putting it in equation (\ref{QspinKin Sz lin 3D}) we find $\delta S_{z}=0$.

The second equation of field (\ref{QspinKin Max lin curl}) gives relations between $\delta B_{x}$, $\delta B_{y}$ and $\delta S_{x}$, $\delta S_{y}$, which are
\begin{equation}\label{QspinKin}\delta B_{x}=4\pi\gamma\int\delta S_{x}d\textbf{p},\end{equation}
and
\begin{equation}\label{QspinKin}\delta B_{y}=4\pi\gamma\int\delta S_{y}d\textbf{p}.\end{equation}
Expressing $\delta S_{x}$ and $\delta S_{y}$ via $\delta B_{x}$ and $\delta B_{y}$ from equations (\ref{QspinKin Sx lin 3D}), (\ref{QspinKin Sy lin 3D}) we come to the following set of algebraic equations
\begin{equation}\label{QspinKin alg Bx By 1}(1-4\pi\gamma \zeta)\delta B_{x}-4\pi\imath\gamma\varsigma\delta B_{y}=0,\end{equation}
and
\begin{equation}\label{QspinKin alg Bx By 2}4\pi\imath\gamma\varsigma\delta B_{x}+(1-4\pi\gamma \zeta)\delta B_{y}=0.\end{equation}
Coefficients $\zeta$ and $\varsigma$ are defined via equilibrium distribution functions $f_{0}(\textbf{p})$ and $S_{0}(\textbf{p})$.

Equations (\ref{QspinKin alg Bx By 1}) and (\ref{QspinKin alg Bx By 2}) have nonzero solution if the determinant of this set of equations equals to zero. This condition gives the following dispersion equation
\begin{equation}\label{QspinKin disp eq gen parallel 3D}1-8\pi\gamma\zeta+16\pi^{2}\gamma^{2}(\zeta^{2}-\varsigma^{2})=0,\end{equation}
where
$$\zeta=-\frac{2\gamma}{\hbar}\Omega_{\gamma}\int \frac{S_{0}(\textbf{p})}{(\omega-\textbf{k}\textbf{p}/m)^{2}-\Omega_{\gamma}^{2}}d\textbf{p}$$
\begin{equation}\label{QspinKin zeta}+\gamma\int\frac{(\omega-\textbf{k}\textbf{p}/m)(\textbf{k}\nabla_{\textbf{p}})f_{0}(\textbf{p})}{(\omega-\textbf{k}\textbf{p}/m)^{2}-\Omega_{\gamma}^{2}}d\textbf{p},\end{equation}
and
$$\varsigma=-\frac{2\gamma}{\hbar}\int\frac{(\omega-\textbf{k}\textbf{p}/m)S_{0}(\textbf{p})}{(\omega-\textbf{k}\textbf{p}/m)^{2}-\Omega_{\gamma}^{2}}d\textbf{p}$$                                                                                                               \begin{equation}\label{QspinKin varsigma}+\gamma\Omega_{\gamma}\int\frac{(\textbf{k}\nabla_{\textbf{p}})f_{0}(\textbf{p})}{(\omega-\textbf{k}\textbf{p}/m)^{2}-\Omega_{\gamma}^{2}}d\textbf{p}.\end{equation}

To get an explicit form of dispersion equation (\ref{QspinKin disp eq gen parallel 3D}) we should present more detailed description of the equilibrium state our system being at. For this we need to give explicit form of the equilibrium distribution functions $f_{0}$ and $S_{0}$.
In our consideration particles are involved in propagating of a one-dimensional perturbation (plane wave).
In this paper we consider magnetized dielectrics at and take same equilibrium state as in case of the perpendicular propagation, see text around formulas (\ref{QspinKin maxwell distrib}) and (\ref{QspinKin eq spin distr func}). Using equilibrium distribution functions (\ref{QspinKin maxwell distrib}) and (\ref{QspinKin eq spin distr func}) to calculate $\zeta$ and $\varsigma$. So, we have result in term of $Z$ function (\ref{QspinKin Z func definition})
$$\zeta=-\frac{2\gamma}{\hbar}\frac{\sigma n_{0}}{2v_{T}k}(Z(\alpha_{+})-Z(\alpha_{-}))$$
\begin{equation}\label{QspinKin}+\gamma\frac{n_{0}}{T}\biggl(1+\frac{1}{2}\alpha_{-}Z(\alpha_{-})+\frac{1}{2}\alpha_{+}Z(\alpha_{+})\biggr),\end{equation}
and
$$\varsigma=\frac{2\gamma}{\hbar}\frac{\sigma n_{0}}{2kv_{T}}(Z(\alpha_{-})+Z(\alpha_{+}))$$
\begin{equation}\label{QspinKin}-\gamma\frac{n_{0}}{2T}(\alpha_{+} Z(\alpha_{+})-\alpha_{-} Z(\alpha_{-})).\end{equation}
Using these result we consider the dispersion equation (\ref{QspinKin disp eq gen parallel 3D}).

In leading order on $\alpha_{\pm}$, under condition $\alpha_{\pm}\gg 1$, equation (\ref{QspinKin disp eq gen parallel 3D}) reappears as
\begin{equation}\label{QspinKin}\alpha_{+}\alpha_{-}+8\pi\gamma B(\alpha_{+}-\alpha_{-})-64\pi^{2}\gamma^{2}B^{2}=0,\end{equation}
where
\begin{equation}\label{QspinKin}B=\frac{2\gamma}{\hbar}\frac{\sigma n_{0}}{2kv_{T}},\end{equation}
and gives the following solution
\begin{equation}\label{QspinKin disp dep parallel alpha big}\omega=\mid \Omega_{\gamma}-\frac{8\pi\gamma^{2}\sigma n_{0}}{\hbar}\mid,\end{equation} where we can introduce equilibrium magnetization $M_{0}=\sigma \gamma n_{0}$. The equilibrium magnetization differs from $\gamma n_{0}$ since we have no full magnetization. Magnetization is caused by external magnetic field, so we can use constitutive equation $M_{0}=\kappa B_{0}$, where $\kappa$ is the ratio of the magnetic susceptibility of the magnetic permeability.
We can represent second term using explicit form of the cyclotron frequency (see text after formula (\ref{QspinKin Sz lin 3D})) then we find
\begin{equation}\label{QspinKin}\omega=\mid \Omega_{\gamma}(1-4\pi\kappa)\mid .\end{equation}
Comparing results obtained in this subsection by means of the quantum kinetics with the quantum hydrodynamic results (see Ref. \cite{Andreev IJMP B 12} formula (24)) we find that they coincide. Contribution of the spin density in the quantum Bohm potential was not considered in Ref. \cite{Andreev IJMP B 12}. Including this contribution \cite{Andreev Asenjo} we get the following formula
\begin{equation}\label{QspinKin}\omega=\mid \Omega_{\gamma}(1-4\pi\kappa)\mid+\frac{\hbar k^{2}}{2m}.\end{equation}
In future we need to find how to get contribution of de-Broglie wave dispersion in quantum kinetics.

Landau damping coefficient corresponding to wave (\ref{QspinKin disp dep parallel alpha big}) appears as
$$\nu=\frac{32\pi^{2}\gamma^{2}\sigma n_{0}\sqrt{\pi}}{\hbar}\frac{(\hbar\Omega_{\gamma}-4\pi\gamma^{2}\sigma n_{0})^{2}}{\hbar\Omega_{\gamma}-8\pi\gamma^{2}\sigma n_{0}}\times$$
$$\times\frac{1}{\hbar\Omega_{\gamma}+4\pi\gamma^{2}\sigma n_{0}-8\pi\gamma^{2}\frac{n_{0}}{T}\hbar\Omega_{\gamma}}\times$$
$$\times\biggl[2B(1+4\pi\gamma A)\biggl(\exp(-\alpha_{0+}^{2})-\exp(-\alpha_{0-}^{2})\biggr)$$
$$-A\biggl(\alpha_{0-}\exp(-\alpha_{0-}^{2})+\alpha_{0+}\exp(-\alpha_{0+}^{2})\biggr)$$
\begin{equation}\label{QspinKin} +16\pi\gamma B^{2}\biggl(\frac{1}{\alpha_{0-}}\exp(-\alpha_{0+}^{2})+\frac{1}{\alpha_{+}}\exp(-\alpha_{0-}^{2})\biggr)\Biggr],\end{equation}
where
\begin{equation}\label{QspinKin}A=\gamma\frac{n_{0}}{T},\end{equation}
\begin{equation}\label{QspinKin}\alpha_{0+}=\frac{2(\hbar\Omega_{\gamma}-4\pi\gamma^{2}\sigma n_{0})}{kv_{T}\hbar},\end{equation}
and
\begin{equation}\label{QspinKin}\alpha_{0-}=-\frac{8\pi\gamma^{2}\sigma n_{0}}{kv_{T}\hbar}.\end{equation}

We have considered limit $\alpha_{\pm}\gg 1$. Below, in this section, we present dispersion dependence and damping coefficient when $\alpha_{+}\gg 1$ and $\alpha_{-}\ll 1$.
To present the dispersion dependence we use the following function
\begin{equation}\label{QspinKin}\Xi=\frac{kv_{T}}{32\pi\gamma B}\frac{1-r^{2}}{1+r^{2}},\end{equation}
where parameter $r$ is defined as
\begin{equation}\label{QspinKin r defin}r^{2}=\frac{4\pi\gamma^{2}n_{0}}{T}.\end{equation}
One can find that $r$ appears to be a rate of two frequencies $r=\lambda/kv_{T}$, here $\lambda^{2}=\frac{4\pi\gamma^{2}n_{0}k^{2}}{m}$.
Finally the dispersion dependence has form of
\begin{equation}\label{QspinKin sol for diff alpha parallel prop} \omega=\sqrt{\biggl(\Omega_{\gamma}+\Xi\biggr)^{2}-\frac{1}{2}\frac{k^{2}v_{T}^{2}}{1+r^{2}}}+\Xi.\end{equation}
At small wave vectors $k\rightarrow 0$ we have $\Xi\rightarrow0$. Consequently, formula (\ref{QspinKin sol for diff alpha parallel prop}) $\omega\simeq\mid\Omega_{\gamma}\mid$. Real part of the dispersion equation solution $\omega=\omega_{R}-\imath\nu$ is presented by formula (\ref{QspinKin sol for diff alpha parallel prop}). Corresponding imaginary part appears as
$$\nu=\frac{kv_{T}}{B}(1+8\pi\gamma A)\sqrt{\pi}\biggl[B+\frac{1}{2}A(\alpha_{-}+\alpha_{+}\exp(-\alpha_{+}^{2}))$$
\begin{equation}\label{QspinKin}-2\pi\gamma\biggl(A^{2}\alpha_{+}\exp(-\alpha_{+}^{2})+2AB+\frac{4}{\alpha_{+}}B^{2}\biggr)\biggr].\end{equation}

Considering spin kinetics we expect to have two characteristic frequencies $\Omega_{\gamma}$ and $kv_{T}$. One came from spin dynamics, another one came from particles distribution in the momentum space (temperature effects). However a third characteristic frequency can be found $\lambda^{2}=\frac{4\pi\gamma^{2}k^{2}n_{0}}{m}$ (see formulas (\ref{QspinKin omega--lambda})) and (\ref{QspinKin r defin}), which reveals an interesting structure if we compare it with the plasma frequency (Langmuir waves) $\omega_{Le}^{2}=\frac{4\pi e^{2}n_{0}}{m}$. Making the following 	 substitute $\gamma k\rightarrow e$ we find coincidence of these formulas. Let us admit that, at hydrodynamic description, frequency $\lambda$ appears when we include the spin-current evolution equation in the set of quantum hydrodynamic equations \cite{Andreev spin current}.

We have considered the kinetic treatment of the spin wave dispersion in three dimensional magnetized dielectrics. In this section we presented dispersion for waves propagating both parallel and perpendicular to the external magnetic field. In the next section we shift our attention to the two dimensional systems of magnetized dielectrics.

\section{V. Dispersion of waves in two dimensional system of spinning particles}

We used the set of nonintegral kinetic equations (\ref{QspinKin kinetic equation gen  classic limit with E and B})
and (\ref{QspinKin kinetic equation gen for spin classic limit with E and B}) coupled with the set of the Maxwell equations (\ref{QspinKin electro stat Max in spin chapter}). Having deal with low dimensional systems of particles we have to use more general integral form of kinetic equations
(\ref{QspinKin kinetic equation gen with spin and int}) and (\ref{QspinKin kinetic equation (spin evol) gen with spin and int with two part F}),
where the electric and magnetic fields are not introduced.

In this section we consider two dimensional system of spinning particles in an external magnetic field directed perpendicular to the plane, where particles are located.

The set of quantum kinetic equations in the self-consistent field approximation can be obtained from equations
(\ref{QspinKin kinetic equation gen with spin and int}) and (\ref{QspinKin kinetic equation (spin evol) gen with spin and int with two part F}) putting there substitutions
(\ref{QspinKin def distribution spin two part function self consist field appr}) and (\ref{QspinKin def distribution N2 self consist field}).
$$\partial_{t}f+\frac{1}{m}\textbf{p}\partial_{\textbf{r}}f+\gamma (\nabla_{\textbf{r}}^{\alpha}B^{\beta}_{ext})\nabla_{\textbf{p}}^{\alpha}S^{\beta}(\textbf{r}, \textbf{p},t)$$
\begin{equation}\label{QKin kinetic equation gen with spin and int Self Consist}-\gamma^{2}
\nabla_{\textbf{p}}^{\alpha}S^{\beta}(\textbf{r},\textbf{p},t) \nabla_{\textbf{r}}^{\alpha}\int  G^{\beta\gamma}(\textbf{r},\textbf{r}') S^{\gamma}(\textbf{r}',\textbf{p}',t) d\textbf{r}' d\textbf{p}' =0,
\end{equation}
In this section we have $\textbf{r}=[x, y]$, $\textbf{k}=[k_{x}, k_{y}]$, $\textbf{p}=[p_{x}, p_{y}]$ since we work in the two dimensional space. However vectors of the spin distribution function $\textbf{S}(\textbf{r},\textbf{p},t)$ and the external magnetic field $\textbf{B}_{ext}(\textbf{r},t)$ have three components $\textbf{S}=[S_{x}, S_{y}, S_{z}]$ and $\textbf{B}_{ext}=[B_{x, ext}, B_{y, ext}, B_{z, ext}]$, since the two dimensional layer is located in the three dimensional space.

In kinetic equation for spinning particles appears the spin distribution function. Therefore, for construction of the closed set of equation describing spinning particles we have to find equation evolution of the spin distribution function. For this goal we differentiate spin distribution function with respect to time, after some calculations we find the kinetic equation for spin distribution function evolution
$$\partial_{t}S^{\alpha}(\textbf{r},\textbf{p},t)+\frac{1}{m}\textbf{p}\partial_{\textbf{r}}S^{\alpha}
+\gamma (\nabla^{\beta} B^{\alpha}_{ext}) \nabla_{\textbf{p}}^{\beta} f(\textbf{r}, \textbf{p},t)$$
$$+\gamma^{2}\nabla_{\textbf{p}}^{\beta} f(\textbf{r},\textbf{p},t)
\nabla_{\textbf{r}}^{\beta}\int G^{\alpha\gamma}(\textbf{r},\textbf{r}')S^{\gamma}(\textbf{r}',\textbf{p}',t) d\textbf{r}'d\textbf{p}'$$
$$-\frac{2\gamma}{\hbar}\varepsilon^{\alpha\beta\gamma}\Biggl(S^{\beta}B^{\gamma}_{ext}+\gamma S^{\beta}(\textbf{r},\textbf{p},t)\times$$
\begin{equation}\label{QspinKin kinetic equation (spin evol) gen with spin and int with two part F Self Consist} \times\int G^{\gamma\delta}(\textbf{r},\textbf{r}')S^{\delta}(\textbf{r}',\textbf{p}',t)d\textbf{r}'d\textbf{p}'\Biggr)=0.\end{equation}

We study evolution of spinning particles which have the magnetic moment. So these particles create an electromagnetic field. Considering nonrelativistic theory we have to use quasi static description of electromagnetic field. Thus, we have to consider the magnetic field created by the magnetic moments. Basic Hamiltonian (\ref{QspinKin Ham spinning part}) does not contain the magnetic field created by the magnetic moments explicitly. Only trace of the internal magnetic field is the Green function of the spin-spin interaction. Despite the fact that the magnetic field satisfies to the Maxwell equations. These equations and this magnetic field do not appear in the general set of the quantum kinetic equations. Nevertheless we introduced the internal magnetic field in the self-consistent field approximation, and this field satisfies to the Maxwell equations (we should admit that if we want to get the Maxwell equations we have to use the correct and full Green function of the spin-spin interaction $G^{\alpha\beta}$). Considering system of particles located in a two dimensional XoY plane in the three dimensional space we get the Dirac delta function $\delta(z)$ in the Maxwell equations. It is not very useful to have the Maxwell equations in such form. So we can keep integral terms describing the spin-spin interaction in the kinetic equations. Thus we will use equations (\ref{QKin kinetic equation gen with spin and int Self Consist}) and (\ref{QspinKin kinetic equation (spin evol) gen with spin and int with two part F Self Consist}) to study dispersion of the collective excitations.

\subsection{A. Some details of dispersion calculation}

Let us extract amplitudes of the distribution function perturbations
\begin{equation}\label{QspinKin}\delta f=F_{A}(\textbf{p})e^{-\imath\omega t+\imath \textbf{k} \textbf{r}},\end{equation}
and
\begin{equation}\label{QspinKin}\delta \textbf{S}=\textbf{S}_{A}(\textbf{p})e^{-\imath\omega t+\imath \textbf{k} \textbf{r}}.\end{equation}

In the linear approximation the set of kinetic equations (\ref{QspinKin kinetic equation (spin evol) gen with spin and int with two part F Self Consist}) have following form
$$-\imath\varpi S_{A}^{x}-\Omega_{\gamma}S_{A}^{y}+\chi\frac{2\gamma^{2}}{\hbar}Q S_{A}^{y}$$
$$+\imath\gamma^{2}(g^{xx}I^{x}+g^{xy}I^{y})\textbf{k}\nabla_{\textbf{p}}f_{0}(\textbf{p})$$
\begin{equation}\label{QspinKin SAx 2D}+\frac{2\gamma^{2}}{\hbar}(g^{yx}I^{x}+g^{yy}I^{y})S_{0}(\textbf{p})=0,\end{equation}
$$-\imath\varpi S_{A}^{y}+\Omega_{\gamma}S_{A}^{x}-\chi\frac{2\gamma^{2}}{\hbar} Q S_{A}^{x}$$
$$+\imath\gamma^{2}(g^{yx}I^{x}+g^{yy}I^{y})\textbf{k}\nabla_{\textbf{p}}f_{0}(\textbf{p})$$
\begin{equation}\label{QspinKin SAy 2D}-\frac{2\gamma^{2}}{\hbar}(g^{xx}I^{x}+g^{xy}I^{y})S_{0}(\textbf{p})=0,\end{equation}
and
\begin{equation}\label{QspinKin SAz 2D}-\imath\varpi S_{A}^{z}+\imath\gamma^{2} g^{zz}I^{z}\textbf{k}\nabla_{\textbf{p}}f_{0}(\textbf{p})=0,\end{equation}
where
\begin{equation}\label{QspinKin}\varpi=\omega-\frac{\textbf{k}\textbf{p}}{m}\end{equation}
is the shifted frequency;
\begin{equation}\label{QspinKin}\textbf{I}=\int \textbf{S}_{A}(\textbf{p}) d\textbf{p}\end{equation}
is an integral of the amplitude of the spin distribution function perturbation over the momentum. It gives us another presentation of oscillating variable.
$I$ is the amplitude of magnetization perturbation;
\begin{equation}\label{QspinKin F transf of G xy}g^{\alpha\beta}=\int G^{\alpha\beta}(\mid \textbf{r}-\textbf{r}'\mid)\exp(\imath \textbf{k}(\textbf{r}'-\textbf{r}))d(\textbf{r}'-\textbf{r})\end{equation}
is the Fourier transformation of the Green function of the spin-spin interaction.
Calculating (\ref{QspinKin F transf of G xy}) we get the following result
\begin{equation}\label{QspinKin F transf of G xy result}g^{\alpha\beta}=\frac{2\pi}{k}(k^{2}\delta^{\alpha\beta}-k^{\alpha}k^{\beta}),\end{equation}
for $\alpha$, $\beta$ equal to $x$ and $y$, $\emph{and}$ $g^{xz}=g^{zx}=g^{yz}=g^{zy}=0$; and
\begin{equation}\label{QspinKin}G^{zz}=-\frac{1}{r^{3}}-\frac{2}{3}\triangle\frac{1}{r}=-\frac{5}{3}\triangle\frac{1}{r},\end{equation}
we remind that $r=\sqrt{x^{2}+y^{2}}$,
so we have
\begin{equation}\label{QspinKin}g^{zz}=\frac{10\pi}{3}k;\end{equation}
Equations (\ref{QspinKin SAx 2D}) and (\ref{QspinKin SAy 2D}) also contain
\begin{equation}\label{QspinKin}\chi=2\pi\frac{5}{3}\int_{r_{min}}^{\infty} \frac{1}{\xi^{2}}d\xi=\frac{5}{3}\frac{2\pi}{r_{min}}\end{equation}
is the integral of $G^{zz}$ over the coordinate space and $r_{min}$ is the minimal distance between particles. For neutral atoms it equals to a diameter of atom.; and
\begin{equation}\label{QspinKin definition of Q}Q=\int S_{0}(\textbf{p})d\textbf{p}\end{equation}
is the integral of the equilibrium spin distribution function over the momentum giving us an equilibrium spin density, which is the equilibrium magnetization $M_{0}$ divided on the Bohr magneton $\gamma$, $Q=M_{0}/\gamma$.

We have two sets of variables related to each other. One of them is the set of amplitudes of the spin distribution functions $S_{A}^{x}$, $S_{A}^{y}$, and $S_{A}^{z}$. The second one is the set of integrals of the spin distribution functions over momentum $I^{x}$, $I^{y}$, and $I^{z}$. Since coefficients in equations (\ref{QspinKin SAx 2D}) and (\ref{QspinKin SAy 2D}) depend on the momentum we can not integrate these equations. However we can express $S_{A}^{x}$, $S_{A}^{y}$, and $S_{A}^{z}$ via $I^{x}$, $I^{y}$, and $I^{z}$. We can integrate obtained dependencies. Integrating formulas for $S_{A}^{x}$ and $S_{A}^{y}$ we have a closed set of algebraic equations for for $I^{x}$ and $I^{y}$, which is
$$I^{x}=\gamma^{2}(g^{xx}I^{x}+g^{xy}I^{y})W_{2}-\imath\frac{2\gamma^{2}}{\hbar}(g^{yx}I^{x}+g^{yy}I^{y})W_{4}$$
\begin{equation}\label{QspinKin eq for Ix} +\Gamma\biggl(\imath\gamma^{2}(g^{yx}I^{x}+g^{yy}I^{y})W_{1}-\frac{2\gamma^{2}}{\hbar}(g^{xx}I^{x}+g^{xy}I^{y})W_{3}\biggr) ,\end{equation}
and
$$I^{y}=-\Gamma\biggl(\imath\gamma^{2}(g^{xx}I^{x}+g^{xy}I^{y})W_{1}+\frac{2\gamma^{2}}{\hbar}(g^{yx}I^{x}+g^{yy}I^{y})W_{3}\biggr)$$
\begin{equation}\label{QspinKin eq for Iy} +\gamma^{2}(g^{yx}I^{x}+g^{yy}I^{y})W_{2}+\imath\frac{2\gamma^{2}}{\hbar}(g^{xx}I^{x}+g^{xy}I^{y})W_{4} ,\end{equation}
where
\begin{equation}\label{QspinKin}\Gamma=\Omega_{\gamma}-\chi\frac{2\gamma^{2}}{\hbar}Q\end{equation}
is the shifted cyclotron frequency,
\begin{equation}\label{QspinKin}W_{1}=\int \frac{\textbf{k}\nabla_{\textbf{p}}f_{0}(\textbf{p})}{\varpi^{2}-\Gamma^{2}}d\textbf{p},\end{equation}
\begin{equation}\label{QspinKin}W_{2}=\int \frac{\varpi \textbf{k}\nabla_{\textbf{p}}f_{0}(\textbf{p})}{\varpi^{2}-\Gamma^{2}}d\textbf{p},\end{equation}
\begin{equation}\label{QspinKin}W_{3}=\int \frac{S_{0}(\textbf{p})}{\varpi^{2}-\Gamma^{2}}d\textbf{p},\end{equation}
and
\begin{equation}\label{QspinKin}W_{4}=\int \frac{\varpi S_{0}(\textbf{p})}{\varpi^{2}-\Gamma^{2}}d\textbf{p}.\end{equation}

We also have following formula for $S_{A}^{z}$
\begin{equation}\label{QspinKin SzIz}S_{A}^{z}=\frac{10\pi\gamma^{2}}{3}kI^{z}
\frac{\textbf{k}\nabla_{\textbf{p}}f_{0}(\textbf{p})}{\omega-\textbf{p}\textbf{k}/m}.\end{equation}
We see that we get independent equations for $I^{x}, I^{y}$ and $I^{z}$. In the next subsection we consider them separately.

\subsection{B. Low dimensional dispersion dependence}

\subsubsection{1. $S_{z}$ evolution}

Integration of equation (\ref{QspinKin SzIz}) over the momentum and reducing $I^{z}$ we get the following dispersion equation
\begin{equation}\label{QspinKin disp eq 2D Sz gen} 1-k\gamma^{2}\cdot\frac{10\pi}{3}\int \frac{\textbf{k}\nabla_{\textbf{p}}f_{0}(\textbf{p})}{\omega-\textbf{p}\textbf{k}/m}d\textbf{p}=0\end{equation}
containing the equilibrium distribution function $f_{0}(\textbf{p})$. In this section, as in previous one, we use the Maxwell distribution function (\ref{QspinKin maxwell distrib}) to describe an equilibrium state. To solve equation (\ref{QspinKin disp eq 2D Sz gen}) and get $\omega(k)$ we have to use an explicit form of the equilibrium distribution function.
\begin{equation}\label{QspinKin gen disp eq Sz 2D} 1-\gamma^{2} k\frac{10\pi}{3}\frac{n_{0}}{T}\biggl(1+\alpha Z(\alpha)\biggr)=0\end{equation}
In leading order on $\alpha$ real part of solution of equation (\ref{QspinKin gen disp eq Sz 2D}) is
\begin{equation}\label{QspinKin Sz 2D zero ord}\omega^{2}_{0,Re}=-\frac{10\pi}{3}\frac{n_{0}}{m}\gamma^{2} k^{3}<0.\end{equation}
In the next order we find contribution of thermal motion in dispersion of the wave
$$\omega^{2}_{Re}=\frac{1}{2}\biggl(-\gamma^{2} k^{3}\frac{10\pi}{3}\frac{n_{0}}{m}$$
\begin{equation}\label{QspinKin Sz 2D 1 ord}+\gamma\sqrt{10k^{3}\frac{n_{0}}{m}} \sqrt{\pi^{2}\gamma^{2} k^{3}\frac{10}{9}\frac{n_{0}}{m}-4\pi v_{T}^{2}k^{2}}\biggr),\end{equation}
which is also negative.

\subsubsection{2. $S_{x}$ and $S_{y}$ evolution}

Nonzero solution of the algebraic equations for $I^{x}$, $I^{y}$ (\ref{QspinKin eq for Ix}), (\ref{QspinKin eq for Iy}) exists if the determinant of this set equals to zero.
It gives the dispersion equation
\begin{equation}\label{QspinKin disp eq SxSy 2D gen}1-2\pi\gamma^{2} k W_{2}+\frac{4\pi\gamma^{2}}{\hbar}k\Gamma W_{3}=0,\end{equation}
where we find no trace of $W_{1}$ and $W_{4}$, since this formula contains  $W_{2}$ and $W_{3}$ only.

Calculating integrals $W_{2}$ and $W_{3}$ via the Maxwell equilibrium distribution function we find
$$1-2\pi\gamma^{2} k \frac{n_{0}}{T}\biggl(1+\frac{1}{2}\alpha_{-}Z(\alpha_{-})+\frac{1}{2}\alpha_{+}Z(\alpha_{+})\biggr)$$
\begin{equation}\label{QspinKin Sx Sy disp general 2D}+\frac{4\pi\gamma^{2}}{\hbar}k\frac{\sigma n_{0}}{2kv_{T}}(Z(\alpha_{+})-Z(\alpha_{-}))=0,\end{equation}
where $\alpha_{-}=(\omega-\Gamma)/kv_{T}$, and $\alpha_{+}=(\omega+\Gamma)/kv_{T}$.

Equations (\ref{QspinKin Sx Sy disp general 3D perp}) and (\ref{QspinKin Sx Sy disp general 2D}) are similar. Replacing $n_{0}(3D)$ by $n_{0}(2D)k/2$ and $\Omega_{\gamma}$ by $\Gamma$ we get from three dimensional case to two dimensional case. This behavior has an analogy with two- and three dimensional Langmuir frequencies differ from each other by the same replacement for the particle concentration. For illustration, let us to present the Langmuir frequencies
\begin{equation}\label{QspinKin Langmuir 3D} \omega_{3D}^{2}=\frac{4\pi e^{2}n_{0}}{m},\end{equation}
and
\begin{equation}\label{QspinKin Langmuir 2D} \omega_{2D}^{2}=\frac{2\pi e^{2}n_{0}k}{m}.\end{equation}
We see linear dependence of the square of frequency on the wave vector module in the two dimensional (\ref{QspinKin Langmuir 2D}), when in the three dimensional case, we find constant frequency depending on the parameters of system (\ref{QspinKin Langmuir 3D}).

At $\alpha_{+}\gg 1$ and $\alpha_{-}\gg 1$
we obtain
\begin{equation}\label{QspinKin SxSy 2D sol}\omega_{R}^{2}=\Gamma\biggl(\Gamma-\frac{4\pi\sigma n_{0}\gamma^{2}k}{\hbar}\biggr),\end{equation}
for real part of the frequency, where the last term contains
$\sigma$, which is proportional to the magnetic susceptibility. In the case of small magnetic susceptibility we find
$$\omega_{R}=\Gamma=\Omega_{\gamma}-\chi\frac{2\gamma^{2}}{\hbar}Q$$
\begin{equation}\label{QspinKin}=\Omega_{\gamma}-\frac{5}{3}\frac{4\pi\gamma^{2}}{r_{min}\hbar}Q,\end{equation}
where $Q$ is defined by formula (\ref{QspinKin definition of Q}). For imaginary part of the frequency we find
$$\nu=\frac{\Gamma}{kv_{T}\omega_{R}}\biggl(\frac{2\pi\sigma n_{0}\gamma^{2}k}{\hbar}\biggr)^{2}\times$$
$$\times\biggl[ \biggl(1+\frac{(\omega_{R}-\Gamma)\hbar}{2\sigma T}\biggr) \exp\biggl(-\frac{(\omega_{R}-\Gamma)^{2}}{k^{2}v_{T}^{2}}\biggr)$$
\begin{equation}\label{QspinKin damping XY 2D}+\biggl(1+\frac{(\omega_{R}+\Gamma)\hbar}{2\sigma T}\biggr)\exp\biggl(-\frac{(\omega_{R}+\Gamma)^{2}}{k^{2}v_{T}^{2}}\biggr)\biggr],\end{equation}
where $\omega_{R}$ is defined by formula (\ref{QspinKin SxSy 2D sol}).

At $\alpha_{+}\gg 1$ and $\alpha_{-}\ll 1$ we get the following solution of equation (\ref{QspinKin disp eq SxSy 2D gen})
$$\omega=\frac{1}{4D}\Biggl(\biggl[k^{2}v_{T}^{2}(1-F)^{2}+16D^{2}\Omega_{\gamma}^{2}-kv_{T}(1-F)$$
\begin{equation}\label{QspinKin perp pr 02 01 2D}-8D kv_{T}[(1-F)\Omega_{\gamma}-Dkv_{T}]\biggr]^{1/2}-kv_{T}(1-F)\Biggr),\end{equation}
where
\begin{equation}F_{2D}=\pi\gamma^{2}k\frac{n_{0}}{T},\end{equation}
and
\begin{equation}D_{2D}=\frac{2\pi\gamma^{2}}{\hbar}k\frac{\sigma n_{0}}{kv_{T}}.\end{equation}
Corresponding imaginary part of frequency $\omega=\omega_{R}-\imath\nu$ is obtained as
\begin{equation}\label{QspinKin SxSy damping 2D} \nu=-\frac{\sqrt{\pi}}{2}kv_{T}\biggl(\sigma+\frac{\hbar(\omega_{R}-\Omega_{\gamma})}{mv_{T}^{2}}\biggr),\end{equation}
where $\omega_{R}$ is defined by formula (\ref{QspinKin perp pr 02 01 2D}). Solution reveals a slow instability of the wave with frequency given by formula (\ref{QspinKin perp pr 02 01 2D}).

As it was expected (see text after formula (\ref{QspinKin Sx Sy disp general 2D})) we have found a lot similarity between dispersion of the spin waves in a two dimensional layer and dispersion of waves propagating perpendicular to external magnetic field in three dimensional medium.

\section{VI. Conclusion}

This paper was written in an attempt to get microscopically proved quantum kinetic theory. We employed the full many-particle wave function governed by the corresponding Pauli equation. We presented a definition of the distribution function in accordance with the classic kinetic theory and its microscopic justification. Our results correspond to the many-particle quantum hydrodynamics, and the many-particle quantum hydrodynamics appears in the full form at applying of the Chapman-Enskog theory to our kinetic equations. In major points our theory coincides with the Wigner's theory, which is the most famous method at the time. But our treatment has differences from Wigner's at description of spinning particles. This fact led us to choosing of the paper topic.

Let us to describe our method itself. New method of the quantum kinetic equation derivation was developed.
This method is the direct generalization of the many-particle quantum hydrodynamics.
In the self-consistent field approximation we got a closed set of two kinetic equations.
It was shown that we need to have two distribution functions for description of spinning particles.
The spin distribution function appears along with the usual distribution function $f(\textbf{r},\textbf{p},t)$.
The spin distribution function is a vector function containing information  about spin direction distribution. We used these equation for spin wave studying.
We considered three- and two-dimensional systems of neutral particles being in an external uniform magnetic field.
We considered two cases for three dimensional system, when waves propagate parallel and perpendicular to the direction of external magnetic field.
In two dimensional case we assumed that an external field is directed perpendicular to the sample, when waves propagate in the plane, where the sample is.
So we have that direction of wave propagation is perpendicular to the external field.

Let us summarize results obtained for wave dispersion.

When we consider wave propagation perpendicular to the external magnetic field we obtained two dispersion relations and we found four wave solutions, each of dispersion equations give two solutions. Dispersion equation (\ref{QspinKin Disp eq Perp prop z via Z}) gives solutions (\ref{QspinKin omega--lambda}) and (\ref{QspinKin Sz 3D small alpha}). Solution (\ref{QspinKin omega--lambda}) reveals negative square of frequency and shows no oscillating behavior. Solution (\ref{QspinKin Sz 3D small alpha}), appearing at $\alpha\ll1$, exists at small enough temperature $T<4\pi\gamma^{2}n_{0}$. Second dispersion equation (\ref{QspinKin Sx Sy disp general 3D perp}) has a solution (\ref{QspinKin perp pr 01 01}) at $\alpha_{\pm}\gg1$ and another solution (\ref{QspinKin perp pr 02 01}) at $\alpha_{+}\gg1$, $\alpha_{-}\ll1$. For all of these wave solution we calculated damping rate which are analogs of the Landau damping of the Langmuir waves in plasma.

Considering propagation of waves parallel to the external magnetic field we got one dispersion equation (\ref{QspinKin disp eq gen parallel 3D}), which we considered in two limits $\alpha_{\pm}\gg1$, and $\alpha_{+}\gg1$, $\alpha_{-}\ll1$. The first case $\alpha_{\pm}\gg1$ gave a solution (\ref{QspinKin disp dep parallel alpha big}) and we have one solution (\ref{QspinKin sol for diff alpha parallel prop}) as well. Damping rate for these waves were calculated either.

In two dimensional case we supposed that the external field was directed perpendicular to the sample. Evolution of $S_{z}$ reveals a solution giving negative square of frequency. So, it support no wave propagation, as it was in three dimensional case. We have wave solutions (\ref{QspinKin SxSy 2D sol}) and (\ref{QspinKin perp pr 02 01 2D}) corresponding to $S_{x}$, $S_{y}$ evolution.

Having it we can trace how dispersion properties change when we go from wave propagation parallel to external field to wave propagation perpendicular to external field.  We found no waves caused by $S_{z}$ evolution in three dimensional systems, at least when wave propagate parallel or perpendicular to the external magnetic field. However, in the case of perpendicular propagation, we got solution (\ref{QspinKin omega--lambda}), whereas we find $\delta S_{z}=0$ at the parallel propagation. Solutions (\ref{QspinKin perp pr 02 01}) and (\ref{QspinKin sol for diff alpha parallel prop}) were obtained at the same conditions, but for wave propagation in different directions. So, they are two limit cases of a general formula appearing at consideration of oblique propagation. We found appearing of thermal effects in the dispersion dependence when we got from parallel propagation to the perpendicular propagation (\ref{QspinKin perp pr 02 01}). In the result we see that the dispersion dependence for the perpendicular propagation is described by the rather large formula (\ref{QspinKin perp pr 02 01}).

We found similarity between three dimensional case for waves propagating perpendicular to the external magnetic field and two dimensional case, since waves propagate perpendicular to external field in both cases.

\section{Appendix}

Considering three dimensional physical space we can represent the Green function of the spin-spin interaction in several equivalent forms. They are
\begin{equation}\label{QspinKin Ap G 1} G^{\alpha\beta}_{pn}=(\partial^{\alpha}_{p}\partial^{\beta}_{p}-\delta^{\alpha\beta}\triangle_{p})(1/r_{pn})\end{equation}
\begin{equation}\label{QspinKin Ap G 2} =4\pi\delta^{\alpha\beta}\delta(\textbf{r}_{pn})+\partial^{\alpha}_{p}\partial^{\beta}_{p}(1/r_{pn})\end{equation}
\begin{equation}\label{QspinKin Ap G 3} =-\frac{\delta^{\alpha\beta}}{r^{3}}+3\frac{r^{\alpha}r^{\beta}}{r^{5}}+\frac{8\pi}{3}\delta^{\alpha\beta}\delta(\textbf{r})\end{equation}
\begin{equation}\label{QspinKin Ap G 4} =-\frac{\delta^{\alpha\beta}}{r^{3}}+3\frac{r^{\alpha}r^{\beta}}{r^{5}}-\frac{2}{3}\delta^{\alpha\beta}\triangle\frac{1}{r}.\end{equation}
Each of them has some benefits. Form (\ref{QspinKin Ap G 1}) allows to get the Fourier transform of $g^{\alpha\beta}$ (see formulas (\ref{QspinKin F transf of G xy}) and (\ref{QspinKin F transf of G xy result})). Formula (\ref{QspinKin Ap G 3}) is most explicit form, which does not contain derivatives and shows result of differentiation only, but it also contains $\delta$ function. Presence of $\delta$ function is not useful when we are going to consider a two dimensional layer in the three dimensional space. So, we believe that form (\ref{QspinKin Ap G 4}) is most useful to go in a two dimensional layer. Being in the two dimensional  layer we should not consider motion in the $z$ direction (we have assumed that the layer is located in XoY plane at $z=0$.). So at this step we replace $\textbf{r}=[x, y, z]$ by the two dimensional coordinate $\textbf{r}=[x, y]$. However we also have that $G^{\alpha\beta}$ is bound to spin vector, which can be directed parallel to Oz axes. Consequently we need to consider $G^{zz}$, $G^{iz}=G^{zi}$, where $i=x$ or $y$, along with $G^{xx}$, $G^{xy}=G^{yx}$, $G^{yy}$. Using formula (\ref{QspinKin Ap G 4}) for the two dimensional case we find $G^{zz}=-1/r^{3}-(2/3)\triangle(1/r)=-(5/3)\triangle(1/r)$.


\begin{acknowledgements}
The author thanks Professor L. S. Kuz'menkov for fruitful discussions.
\end{acknowledgements}

\end{document}